 \documentclass[final,5p,times,twocolumn]{elsarticle}

\usepackage{amssymb}
\usepackage{amsmath}
\usepackage{url}
\usepackage{siunitx}

\usepackage{soul}
\usepackage[normalem]{ulem}

\journal{Nuclear Physics A}

\begin{document}

\begin{frontmatter}


\title{Optical studies of scintillation detectors for precision beta-energy measurements}

\author[KUL]{S.~Vanlangendonck}
\author[SCK]{D.~Atanasov}
\author[LP2iB]{B.~Blank}
\author[LPC]{X.~Fléchard}
\author[LPC]{M.~Kanafani}
\author[LP2iB]{S.~Lecanuet}
\author[LPC,FRIB,IRL]{O.~Naviliat-Cuncic}
\author[KUL]{N.~Severijns}
\author[LP2iB]{M.~Versteegen}

\affiliation[KUL]{organization={KU Leuven, Instituut voor Kern- en Stralingsfysica},
            addressline={Celestijnenlaan 200D}, 
            city={Leuven},
            postcode={B-3001},
            country={Belgium}}
\affiliation[SCK]{organization={Belgian Nuclear Research Centre SCK-CEN}, postcode={2400}, city={Mol}, country={Belgium}}
\affiliation[LP2iB]{organization={LP2i Bordeaux, UMR5797, Université de Bordeaux}, postcode={F-33170}, city={Gradignan}, country={France}}
\affiliation[LPC]{organization={Université de Caen Normandie, ENSICAEN, UNICAEN, CNRS/IN2P3, LPC Caen}, postcode={F-14000}, city={Caen}, country={France}}
\affiliation[FRIB]{organization={Facility for Rare Isotope Beams and Department of Physics and Astronomy, Michigan State University}, postcode={48824 MI}, city={East Lansing}, country={USA}}
\affiliation[IRL]{organization={International Research Laboratory for Nuclear Physics and Nuclear Astrophysics, CNRS-MSU}, postcode={48824 MI}, city={East Lansing}, country={USA}}

\begin{abstract}
This work presents optical calculations and simulations for scintillation detectors used in precision measurements of beta-particle energy spectra. Particular attention is given to Cherenkov photons and the impact of the light detection efficiency in the detector ensemble. We present an approach to estimate this light detection efficiency from the measured energy resolution of a detector. This is essential for the inclusion of scintillation photons in optical tracking Monte-Carlo simulations. A method to account for possible saturation and cross-talk of silicon photo-multipliers is also discussed. 
The impact of these effects is quantified in terms of systematic shifts in the extraction of the Fierz interference term from measurements of beta-energy spectra using scintillator detectors.
\end{abstract}

\end{frontmatter}

\section{Introduction}
\label{sec:introduction}

The study of energy spectra of beta particles is commonly used to address several fundamental questions in nuclear physics, including radionuclide metrology \cite{Kossert2022}, nuclear structure properties \cite{George2014, Haaranen2017, Paulsen2024}, as well as searches for exotic
interactions in the weak interaction \cite{Glick2017, Cirgiliano2019, Hughes2019, Byron2023, DeKeukeleere2024}.
In the past decade, much effort has been dedicated to precision measurements of beta-energy spectra because these provide a direct determination of the Fierz interference term, which is linearly sensitive to the small exotic couplings constants \cite{Jackson1957b, Naviliat2013}. 
Current techniques include the use of a light-gas wire chamber with a plastic scintillator detector \cite{DeKeukeleere24b}, calorimetry techniques using implanted radioactive beams on inorganic scintillators \cite{Naviliat2016, Kanafani2025}, the implementation of cyclotron radiation electron spectroscopy \cite{Byron2023} or the use of a radioactive source located between two detectors in a strong magnetic field \cite{Vanlangendonck2023}.  

Past studies in neutron decay \cite{Hickerson2017} and more recently in the decay of $^{114}$In \cite{DeKeukeleere2024, Vanlangendonck2023} showed that the detector nonlinearity is an important systematic effect. This can potentially also be important in the decay of $^{6}$He \cite{Kanafani2023PhD}.
To estimate this effect for techniques implementing scintillation detectors, we have studied three potential sources of nonlinearities in the light emission and collection mechanisms, i) the production of scintillation light (Sec.~\ref{sec:Birks}); ii) the often-overlooked Cherenkov radiation in a scintillator (Sec.~\ref{sec:Cherenkov}); and iii) the light detection by the scintillator readout (Sec.~\ref{sec:readout}). 
To quantify the magnitude of these effects for specific measurements, the adopted geometries are those of the InESS@WISArD \cite{Vanlangendonck2023} and bSTILED \cite{Kanafani2022, Kanafani2023} setups, which are presented in Sec.~\ref{sec:examples}. 
For these geometries we additionally studied the effect of the light collection efficiency. These studies rely, to a large extent, on Monte Carlo (MC) simulations tailored to the experimental setup that incorporate optical tracking of scintillation photons. Further details on these simulations are provided in Sec.~\ref{sec:examples}. 
Finally, in Sec.~\ref{sec:MCfits}, the impact of all the effects is quantified using another MC routine to generate the beta-particle energy spectrum and extract the Fierz interference term when including or omitting the relevant effect. 

\section{Sources of nonlinearities in scintillators}
\label{sec:scintillation}

The three sources of possible nonlinearities discussed here are: the production of scintillation light, the generation of Cherenkov radiation, and the scintillator readout. The physical and technical aspects of each of these sources of nonlinearities are analyzed below.

\subsection{Production of scintillation light and quenching} 
\label{sec:Birks}

When ionizing radiation interacts with a scintillating material, it excites the atoms or molecules within the scintillator. As these excited states decay, they emit photons, resulting in scintillation light. However, when the ionization density along the track is high, the scintillation light output is said to be quenched due to increased non-radiative energy loss mechanisms leading to a reduced scintillation efficiency.

The light production of a particle along its track, $\mathrm{d}L/\mathrm{d}x$, is related to the particle energy loss in the scintillator, $\mathrm{d}E/\mathrm{d}x$, through an empirical relation first suggested by Birks in the form \cite{Birks1951},
\begin{equation} \label{eq:Birks}
    \frac{\mathrm{d}L}{\mathrm{d}x} = S \frac{\mathrm{d}E/\mathrm{d}x}{1 + kB (\mathrm{d}E/\mathrm{d}x)}\,,
\end{equation}
where $S$ is the scintillation efficiency without quenching and the product $kB$ is a single empirically determined parameter which depends on the particle type.

In recent years, significant progress has been made in describing light quenching from first principles, by focusing on the transport of the excitation density within the scintillation material. A useful concept to explain electron-hole recombination and understand the experimental electron response curves is the Onsager radius \cite{Payne2011}. It allows the characterization of materials into different classes which are mainly defined by the host material.
For ions, the best agreement is obtained by solving the Blanc equation \cite{Christensen2018}; the Birks model being only a simplified solution of the latter. This approach allows the calculation of the light quenching in organic plastic scintillators without experimental input but only using the decay time, the density, and the light yield of the scintillation material. 
Despite these significant progresses in the description of quenching from first-principles, measurements remain essential to reach the required level of precision in beta-decay studies.


Experimental light quenching data can be obtained with several approaches, and we only list
here a few examples. For protons or heavier ions, the quenching factor is often determined by comparing the response of the quenched signal in an organic scintillator to the unquenched signal measured with an ionization chamber \cite{Christensen2018, Torrisi2000}. For electrons, the response can be determined using the Compton coincidence technique \cite{Choong2008}. In this technique, a collimated, mono-energetic gamma source illuminates a scintillator and both, the light output produced by the Compton electron, and the outgoing Compton-scattered $\gamma$-ray are detected. Since the energy of the outgoing $\gamma$-ray is determined by the scattering angle, the detection of a $\gamma$-ray at a ﬁxed angle also implies a ﬁxed energy deposited in the scintillator. A measurement of the energy deposition versus light output is then obtained by detecting $\gamma$-rays at different scattering angles.
Taking, for example, a polyvinyltoluene (PVT) scintillator NE-102, the value for $kB$ has been experimentally determined by several authors. For electrons, the published values for $kB$
range from $0.123$ to $0.143$~mm/MeV, with an outlier of $0.202$~mm/MeV \cite{Badhwar1967, Craun1970, Torrisi2000, Zhang2015}.

\begin{figure}[!htb]
    \centering
    \includegraphics[width=\linewidth]{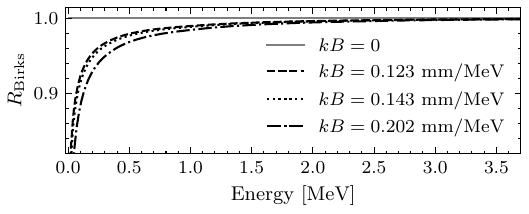}
    \caption{Nonlinearity due to light quenching as described by Eq.~\eqref{eq:Birks}
    using the literature values for a PVT scintillator of the type NE-102.}
    \label{fig:EffectBirksLiterature}
\end{figure}

Figure~\ref{fig:EffectBirksLiterature} illustrates the resulting nonlinearity, defined as 
\begin{equation}
    R_{\rm Birks}=\frac{L(E)}{SE}\,,
\end{equation}
where the light output $L(E)$ is calculated using Eq.~\eqref{eq:Birks}.
As seen, the nonlinearity induced by quenching is small for fast electrons, that is, when $dE/dx$ is sufficiently small, but becomes significant for low-energy electrons.

\subsection{Nonlinearity induced by Cherenkov photons}
\label{sec:Cherenkov}

If the speed of a particle exceeds the speed of light in the optical material, the depolarization waves interfere constructively resulting in so-called Cherenkov radiation. The optical photons produced and detected in a scintillator originate therefore from scintillation as well as the Cherenkov process. Unlike the former, the number of Cherenkov photons does not scale, even approximately, linearly with the energy deposition. 
As a consequence, Cherenkov radiation is a source of detector nonlinearity that could be of importance when high-precision is required. Moreover, the effect of Cherenkov radiation in a scintillator is often overlooked.

The total number of Cherenkov photons emitted by an electron along its track,
in the spectral window delimited by $\lambda_{\rm min}$ and $\lambda_{\rm max}$,
is given by the classical formula \cite{FrankTamm1937}
\begin{equation} \label{eq:dNmm_wavelength}
\frac{\mathrm{d} N}{\mathrm{d} L}=2 \pi \alpha \left(\frac{1}{\lambda_{\rm min}}-\frac{1}{\lambda_{\rm max}}\right) \sin ^2 \theta_C\,,
\end{equation}
with $\alpha$ the fine-structure constant, and $\sin^2 \theta_C = 1 - 1 / (\beta_e n)^2$ with $n$ the index of refraction of the optical medium and $\beta_e = v/c = pc/E_e$ the electron velocity.
To get the total number of Cherenkov photons that are emitted in addition to the scintillation photons, Eq.~\eqref{eq:dNmm_wavelength} should be integrated over the path length of the electron in the scintillator,
\begin{equation}
\label{eq:theory_cherenkov_number}
    N_{\rm Cer} (E)  = 2 \pi \alpha \left(\frac{1}{\lambda_{\rm min}}-\frac{1}{\lambda_{\rm max}}\right) \int_{E_{th}}^{E} \frac{1}{\rho} \frac{\sin^2 \theta_C}{S(E)} \mathrm{d}E, 
\end{equation}
where $\lambda_{\rm min}$ and $\lambda_{\rm max}$ are defined by the spectral sensitivity of the readout device, $E$ is the initial electron kinetic energy, $E_{th}$ the threshold energy to generate Cherenkov radiation, $\rho$ the density of the detector material, and $S(E) = -(1/\rho)(dE/dx)$. The latter can for instance be found in the ESTAR tables \cite{estar}.
A previous study reported the number of Cherenkov photons emitted as a function of electron energy in different materials~\cite{Sowerby1971} (cited by Knoll in Fig.~19.2 \cite{Knoll2010}). We checked these calculations with identical material properties and found them to overestimate the number of Cherenkov photons by more than a factor 2.

We characterize the nonlinearities induced by Cherenkov photons with the ratio
\begin{equation}
\label{eq:Cherenkov_nonlinearity}
R_{\rm Cer} = \frac{N_{\rm Cer} + N_{\rm Scint}}{N_{\rm Scint}}\,,
\end{equation}
where $N_{\rm Cer}$ ($N_{\rm Scint}$) is the number of produced or detected Cherenkov (scintillation) photons. 
Deducing the nonlinearity from the total number of produced Cherenkov photons, as calculated using Eq.~\eqref{eq:theory_cherenkov_number}, constitutes only a first approximation since it does not take into account differences in the light detection efficiencies (LDE), $\varepsilon_{\mathrm{LD}}$.
This efficiency is the product between the light collection efficiency (LCE) and the photo-detection efficiency (PDE): $\varepsilon_{\rm LD} = \varepsilon_{\rm LC} \varepsilon_{\rm PD}$.
The LCE is the ratio between the number of photons hitting the surface of the readout device and the total number of generated photons, whereas the PDE depends on the readout device.

\begin{figure}
    \centering
    \includegraphics[width=\linewidth]{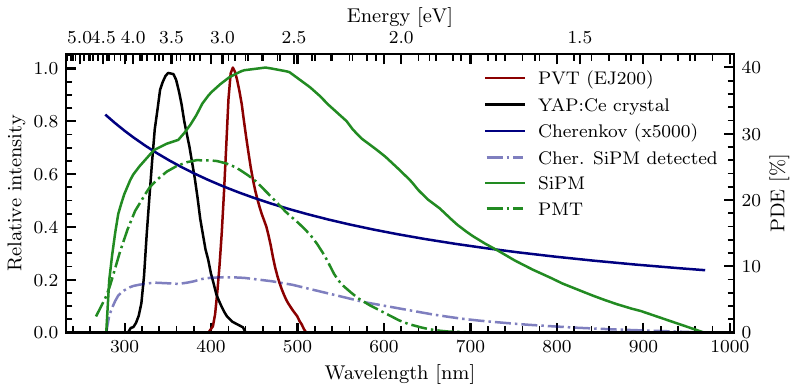} 
    \caption{The scintillation emission spectrum for PVT EJ200 (red) and YAP:Ce (black) compared with the Cherenkov spectrum (blue) and the PDE of a SiPM-Hamamatsu S13360-6050CS (solid green) and a PMT Hamamatsu R7723 (dash-dotted green). The blue dash-dotted line is the product of the Cherenkov spectrum and the SiPM PDE. Note the second axis for the PDE.}
    \label{fig:ES_overview}
\end{figure}

A detailed discussion of the LCE, for dedicated scintillator geometries, is presented in Sec.~\ref{sec:examples}. Here we assume an equal LCE for both types of photons and focus on the varying PDE within the range of wavelengths being considered.

\begin{figure}
    \centering
    \includegraphics[width=\linewidth]{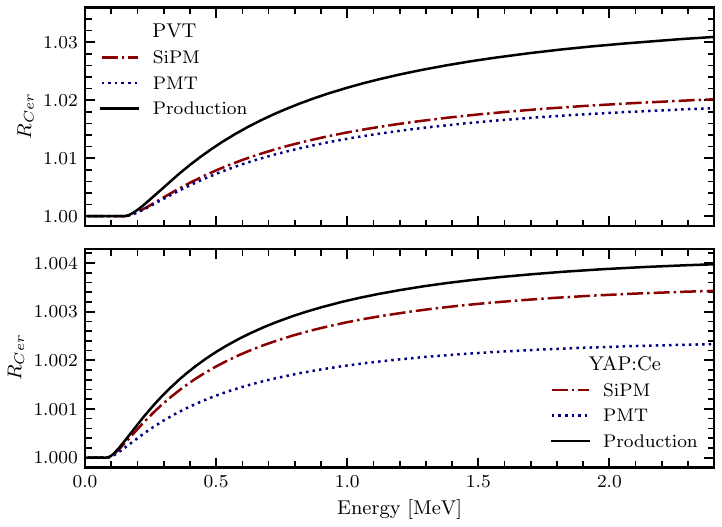} 
    \caption{Nonlinearities deduced from Eq.~\eqref{eq:Cherenkov_nonlinearity}, induced by Cherenkov photons in the $278.8-970$~nm range. Top: for a PVT EJ200 scintillator, with $n=1.58$, $\rho = 1.023~g/cm^3$, and threshold energy $E_{\rm th} = 149.0$~keV.
    Bottom: for a YAP:Ce scintillator with $n=1.95$, $\rho = 5.37~g/cm^3$, and $E_{\rm th} =84.2$~keV. Note the different y-scale for both materials. See text for details. }
    \label{fig:Cherenkov_nonlinearity_example}
\end{figure}

The spectral sensitivity of the readout device generally matches the scintillation spectrum, as illustrated in Fig.~\ref{fig:ES_overview}. As a result, the average PDE for Cherenkov photons is smaller than for scintillation photons. This is also shown in Table~\ref{tab:PDE} which lists the average detection efficiency for Cherenkov photons compared to scintillation photons for a photomultiplier tube (PMT) Hamamatsu R7723 and a silicon photomultiplier (SiPM) Hamamatsu S13360-6050CS. 
As an illustration of these considerations, the nonlinearities induced in PVT (EJ200) \cite{PlasticScintillator} and YAP:Ce \cite{Crytur_YAP} scintillators are shown in Fig.~\ref{fig:Cherenkov_nonlinearity_example}. The solid lines correspond to the number of produced photons, Eq.~\eqref{eq:theory_cherenkov_number}, whereas the dotted and dash-dotted lines to detected photons using those PMT or SiPM, respectively.

\begin{table}[!htb]
    \centering
    \caption{Average PDE ($\varepsilon_{\rm PD}$) for Cherenkov photons, with wavelengths between $278.8$ and $970$~nm, which hit a PMT Hamamatsu R7723 or a SiPM Hamamatsu S13360-6050CS compared to the weighted average PDE over the scintillation spectrum for PVT (EJ200) and YAP:Ce scintillators.}
    \begin{tabular}{l l c c}
    \hline \hline
      &   & PMT & SiPM  \\ 
    \hline
    Cherenkov & & 15.8\% & 25.5\% \\
    Scintillation & PVT  & 23.7\% & 39.1\% \\
     & YAP:Ce  & 24.3\% & 29.6\% \\
    \hline \hline
    \end{tabular}
    \label{tab:PDE}
\end{table}

For the YAP:Ce, the nonlinearity is an order of magnitude smaller, which is partly due to its higher scintillation light yield ($25$ photons/keV, compared to $10$ photons/keV for the PVT) but is mainly due to its higher density, $\rho=5.37$~g/cm$^3$.
The largest nonlinearities are observed for produced (or generated) Cherenkov and scintillation photons, whereas the nonlinearities for detected photons are smaller due to the different average efficiencies of the PMT and SiPM, for both Cherenkov and scintillation photons (Table \ref{tab:PDE}).
These differences in the average detection efficiency reduce the induced nonlinearity by up to a factor 2. 

\subsection{Scintillator readout}
\label{sec:readout}

The final step towards a determination of the energy deposited in a scintillator is the readout device, which is expected to have a linear response. 
A PMT fulfills this requirement since it shows no intrinsic nonlinearity for events where the signal amplitude is well-below charge saturation effects and in the absence of pile-up. 
In contrast, the SiPM response has an intrinsic energy dependence because  of the finite number of pixels working in Geiger mode. The expected number of pixels fired in the SiPM is determined by the product of the number of scintillation photons and the LDE. When the number of pixels fired becomes large compared to the total number of pixels, the response will be quenched due to saturation. Note that this saturation can be strongly enhanced when the SiPM is not irradiated uniformly. Conversely, for a small number of pixels fired, the response can be augmented due to after-pulsing and (optical) cross-talk \cite{VanDam2010}. To estimate this effect, it is thus crucial to know the number of pixels that are expected to be triggered in the energy region of interest.
 
An approximate, empirical input-output transfer function to calculate the average number of pixels triggered, $N_{\rm pix}$, is given by \cite{Renker2009}
\begin{equation}
\label{eq:NmbPixelsSiPM}
\begin{aligned}
    N_{\rm pix} & = N_{\rm tot} \left( 1 - e^{- N_{\rm det}/N_{\rm tot}} \right)\,,
\end{aligned}
\end{equation}
with $N_{\rm tot}$ the total number of pixels on the SiPM and $N_{\rm det} = N_{\rm col} \cdot \varepsilon_{\rm PD}$ the number of photons expected to be detected given by the product of the number of photons hitting the SiPM, $N_{\rm col}$, and the PDE.
Although useful as a first estimate, the properties of the SiPM studied in Ref.~\cite{Renker2009} might not resemble the ones in a specific setup. Moreover, this transfer function provides no information on the peak broadening. Using combinatorics, it is possible to calculate the probability to trigger a certain number of pixels given the number of initial photons.
We elaborate here upon an unpublished approach developed during the analysis of another precision measurement in beta decay \cite{Atanasov2023, AraujoEscalona2020}. The probability that a number of pixels triggers is given by
\begin{equation}
\label{eq:SiPM_nonlinearity_distribution}
P(N_{\rm pix})=\frac{C(N_{\rm tot}, N_{\rm pix}) \times \omega(N_{\rm pix}, N_{\rm det}-N_{\rm pix})}{\omega(N_{\rm tot}, N_{\rm det})}
\end{equation}
where 
$ C(n, k) = n !/((n-k) ! k !)$ are binomial coefficients without repetitions,
and $\omega(n, k) = (n+k-1) !/((n-1) ! k !)$ are coefficients including repetitions.
Calculating this probability for different numbers of pixels, $N_{\rm pix}$, close to the expected value, results in the wanted probability distribution. An additional benefit is that this model allows studying the effect of the SiPM response on the energy resolution \cite{Vanlangendonck2023}.  

Besides the nonlinearity due to saturation, cross-talk is a second important contribution to the SiPM response function. Different models, with increasing levels of complexity, have been published \cite{Vinogradov2012, Gallego2013} and the required level of complexity mainly depends on the cross-talk probability of the SiPM under study. For modern SiPMs, this probability is usually limited to a few percent \cite{HamamatsuTechNotes} and it is therefore sufficient to use a Binomial distribution.

\begin{figure}
    \centering
    \includegraphics[width=\linewidth]{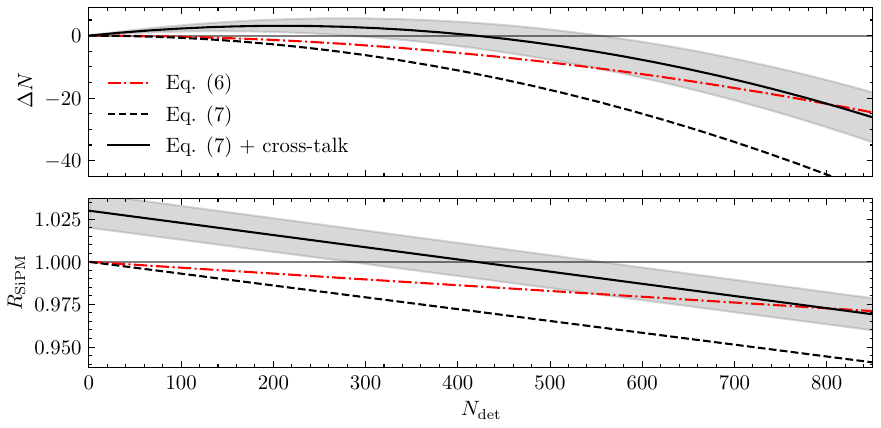}
    \caption{Difference (top) and ratio (bottom) between the average number of pixels triggered, $N_{\rm pix}$, calculated using Eq.~\eqref{eq:NmbPixelsSiPM} or Eq.~\eqref{eq:SiPM_nonlinearity_distribution}, and the number of photons expected to be detected in absence of saturation and cross-talk, $N_{\rm det}$.
    The gray bands illustrate a variation by $\pm 1\%$ of the cross-talk probability from its nominal value of 3\%.} 
    \label{fig:SiPM_non_linearity}
\end{figure}

The nonlinearity induced by saturation and cross-talk can be characterized by the difference $\Delta N = N_{\rm pix} - N_{\rm det}$ and by the ratio
\begin{equation}
\label{eq:SiPM_nonlinearity}
R_{\rm SiPM} = \frac{N_{\rm pix} }{N_{\rm det}}\,. 
\end{equation}
Figure~\ref{fig:SiPM_non_linearity} shows $\Delta N$ and $R_{\mathrm{SiPM}}$ for a SiPM with $N_{\rm tot}=14~400$ pixels and where $N_{\rm pix}$ is calculated using either Eq.~\eqref{eq:NmbPixelsSiPM} or the mean from Eq.~\eqref{eq:SiPM_nonlinearity_distribution}. Without taking cross-talk into account, the calculated deviation from linearity using Eq.~\eqref{eq:SiPM_nonlinearity_distribution} is consistently larger than using Eq.~\eqref{eq:NmbPixelsSiPM}. Adding a realistic cross-talk probability of $3\%$ \cite{HamamatsuTechNotes}, the difference becomes smaller and this model predicts the SiPM response $R_{\mathrm{SiPM}}$ to be larger than unity for small pixel numbers, where the effect of saturation is small \cite{VanDam2010}.

So far, the description of the SiPM response assumed an equal output for every pixel. However, gain variations are possible for experiments with high counting rates or for scintillators with relatively long decay times. 
After a pixel is triggered, the overvoltage is temporarily quenched. The time required to restore $100\%$ of the gain is called the recovery time. 
The recovery time depends on the total size of the SiPM and the photosensitive area. For a pixel pitch of \SI{50}{\micro m} the typical recovery time is $50-100$~ns \cite{HamamatsuTechNotes}. 
To accurately take this time-dependent gain into account, additional characterization measurements are needed. The effect can, for example, be measured by studying the pulse height when using incident light of different frequencies \cite{HamamatsuTechNotes}.
An accurate model for these gain variations includes experimental-specific information, like the temporal characteristics of the scintillation material and the SiPM as well as information on its overvoltage and temperature~\cite{VanDam2010}. It is therefore beyond the scope of the present work to further quantify this effect and we will continue to assume a constant and identical gain for every pixel.

\section{Application to specific experiments}
\label{sec:examples}

To quantify the impact of these nonlinearities to precision
measurements of beta-energy spectra, their effect was evaluated
for the setups of two ongoing experiments, whose acronyms are InESS@WISArD \cite{Vanlangendonck2023} and bSTILED \cite{Kanafani2022, Kanafani2023}. 
To limit the effect of partial energy loss due to electron backscattering, which would otherwise be the dominant systematic effect in such measurements, both experiments implement a $4\pi$~sr solid angle detection geometry. However, they use different types of scintillation detectors and readout devices. As a result, the impact of the nonlinearity effects discussed above are rather different.

Some effects can only be studied in detail using optical tracking or MC simulations. Here, we discuss how this can be performed using the Geant4 toolkit (v4.11.0.0) \cite{Agostinelli2003, Allison2006, Allison2016}. A summary of the important parameters in the optical models used in Geant4 is presented in \ref{app:simulations_G4}.

\subsection{InESS@WISArD}

\begin{figure}
    \centering
    \includegraphics[width=\linewidth]{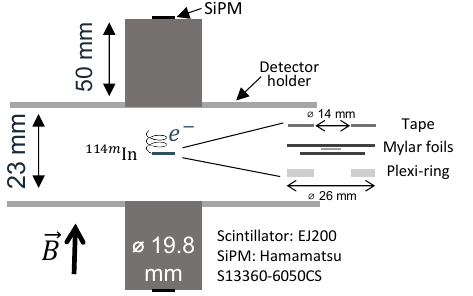}
    \caption{Scheme of the detection setup for the InESS@WISArD proof-of-principle experiment, with a zoom around the $^{114m}$In radioactive source. The set-up consists of two plastic scintillators (grey) in a symmetrical configuration, located at $11.5$~mm from the source in the center. The $6\times6$ mm$^2$ SiPM are shown in black.}
    \label{fig:schematic_setup_InESS}
\end{figure}

The setup of the InESS@WISArD (Indium Energy Spectrum Shape) experiment is schematically shown in Fig.~\ref{fig:schematic_setup_InESS}.
This experiment aims at a precision measurement of the energy spectrum of beta particles from $^{114m}$In decay. Two beta-particle detectors are installed face-to-face in the strong magnetic field of the superconducting solenoid of the WISArD setup  \cite{Atanasov2023, AraujoEscalona2020} at ISOLDE/CERN. The magnetic field, of up to $9$~T, strongly confines the trajectories of beta particles to the sensitive region of the particle detectors.
The $4\pi$~sr solid angle results here from the fact that backscattered beta particles are not lost but guided towards the opposite detector.
This makes efficient coincidence detection vital for the identification of such events. The proof-of-principle experiment \cite{Vanlangendonck2023} used unpolished PVT EJ200 scintillators \cite{PlasticScintillator}. The detectors were $50$~mm long with a diameter of $19.8$~mm and were wrapped with two layers of Teflon tape to improve light collection. Because of the high magnetic field, the light was collected using a $6\times6$ mm$^2$ SiPM S13360-6050CS from Hamamatsu \cite{HamamatsuTechNotes}.
In practice, only a fraction of the scintillation photons are detected. We describe here how this LCE can be experimentally estimated and how this effect can be included to improve the accuracy of simulations. 

\subsubsection{Light detection efficiency}
\label{sec:light_collection_eff_InESS}

For a scintillator, the energy resolution is typically limited by the statistical fluctuations of the number of scintillation photons produced and subsequently detected at a given energy. When using a PMT, this fluctuation arises from variations in the number of photoelectrons emitted by the photocathode. For a SiPM, it is the variation in the number of pixels that are triggered. In both cases, the LDE can be inferred from the measured energy resolution.

The energy resolution is typically determined using several mono-energetic peaks, with
a kinetic energy $E$.
The ratios $R = \Delta E_{\rm FWHM}/E$ are then fitted with a function of the form,
\begin{equation} \label{eq:res_exp}
    R = \frac{\Delta E_{\rm FWHM}}{E} = \sqrt{ \alpha^2 + \left( \frac{\beta}{\sqrt{E}}\right)^2 + \left(\frac{\gamma}{E}\right)^2}
\end{equation}
where $\alpha$ is associated to geometrical effects, $\beta$ is due to statistical fluctuations, and $\gamma$ is a constant contribution due to electronic noise.
Assuming that the second term dominates, which is driven by the mean number of detected photons, we have $R \approx \beta/\sqrt{E} \approx 2.35/\sqrt{N}$, with $N = \varepsilon_{\rm LD} E Y $, where $N$ is the number of photoelectrons produced in the photocathode of a PMT or the number of pixels triggered on a SiPM\footnote{For SiPMs this is only valid when the cross-talk probability is sufficiently small. The contribution of cross-talk can then be neglected.}, and $Y$ is the scintillation light yield per unit deposited energy.
Inverting the equation we get an estimate of the LDE,
\begin{equation}
\label{eq:res_counts}
   \varepsilon_{\rm LD} \approx \frac{1}{Y} \left(\frac{2.35}{\beta}\right)^2 
\end{equation}
As indicated in Sec.~\ref{sec:Cherenkov},
this efficiency is the product between the LCE and the PDE (Fig.~\ref{fig:ES_overview}). 

\begin{figure}
    \centering
    \includegraphics[width=\linewidth]{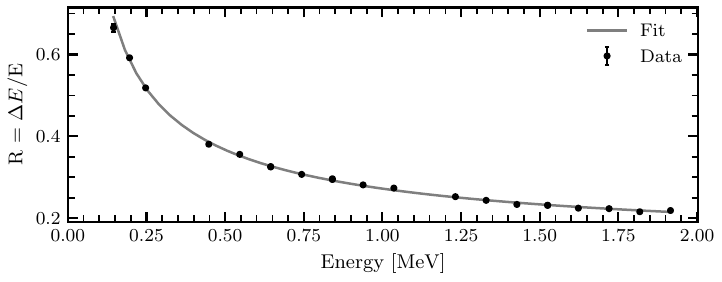}
    \caption{Energy resolution of the InESS detection assembly as determined using mono-energetic electrons from the spectrometer at LP2i in Bordeaux \cite{Marquet2015} and the fit using Eq.~\eqref{eq:res_exp}. The uncertainty is smaller than the markers for most data points.}
    \label{fig:resolution_InESS_Bordeaux}
\end{figure}


The energy resolution of the InESS detection assembly has been characterized in about $20$ measurements with mono-energetic electrons, with energies between $0.25$ and $2.0$ MeV provided by the spectrometer at LP2i in Bordeaux \cite{Marquet2015}. Before determining the energy resolution, the electron energy is corrected for energy loss in the trigger module, which is a \SI{130}{\micro m} thick plastic scintillator wrapped in aluminized mylar foil, and for light quenching using $kB = 0.123 $ mm/MeV (Fig.~\ref{fig:EffectBirksLiterature}). The data and the corresponding fit with Eq.~\eqref{eq:res_exp} are shown in Fig.~\ref{fig:resolution_InESS_Bordeaux}. The resulting fit parameters are $\alpha=0.130(2)$, $\beta=0.2359(17)$~MeV$^{1/2}$, and $\gamma=0.041(2)$~MeV. Using Eq.~\eqref{eq:res_counts}, this resulted in an estimated LDE of only $0.953(13)\%$. The very low LDE in this arrangement originates mainly from the large mismatch between the area of the SiPM and that of the scintillator surface as the SiPM covers only $11.7 \%$ of the surface.  

\begin{figure}
    \centering
    \includegraphics[width=\linewidth]{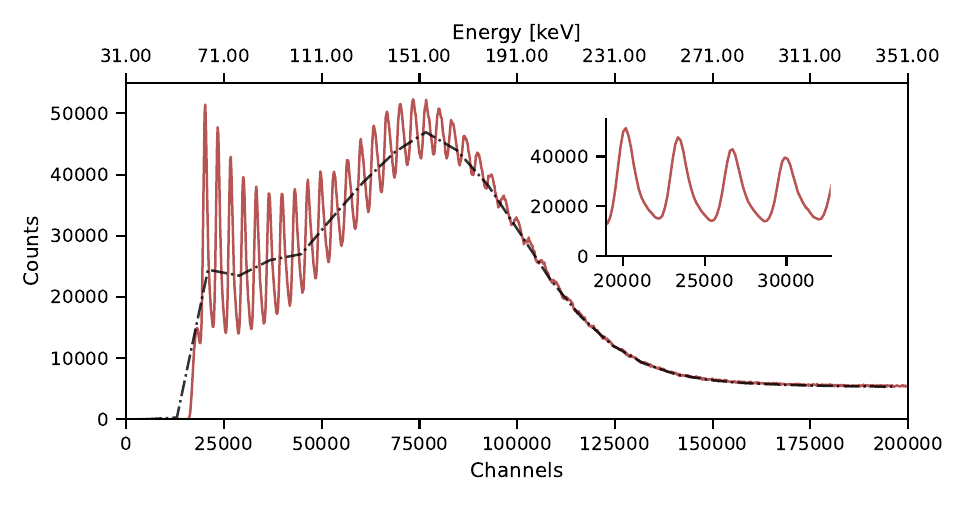}
    \caption{The low energy part of the experimental energy spectrum of electrons from $^{114m}$In decay. When using a fine binning (red), the SiPM single pixel behavior is seen. For comparison, the spectrum with a larger binning (black) is normalized at channel $200~000$. 
    The inset shows a zoom on the first peaks. The tails to the right are attributed to delayed cross-talk and after-pulsing. } 
    \label{fig:few_pixel_peaks_SiPM}
\end{figure}

When collecting the optical photons with a SiPM in a sufficiently stable system, the discrete behavior due to the SiPM pixels can be observed in the low energy part of the energy spectrum, as shown in Fig.~\ref{fig:few_pixel_peaks_SiPM}. The pixel multiplicity at low-energy can then provide another way to estimate the LDE. Each peak corresponds to a given multiplicity, $M$, of pixels triggered by the scintillation photons. As all SiPM pixels have a constant and identical gain, the different peaks are equidistant and the separation between the peaks provides a measurement for this gain. Moreover, the width of these peaks increases due to the statistical broadening (proportional to $\sqrt{M}$) induced by the increasing number of pixels that trigger. The asymmetric tail of the peaks (inset of Fig.~\ref{fig:few_pixel_peaks_SiPM}) is attributed to delayed cross-talk and after-pulsing \cite{Chmill2017}.

These multiple pixel peaks provide additional means to estimate the LDE by determining the deposited energy per pixel. 
First, a multiple-peak fit on the first few peaks allows the determination of the interpeak separation in channels. Next, a fit of the energy deposition spectrum using a histogram with larger binning, as shown in black in Fig.~\ref{fig:few_pixel_peaks_SiPM}, is used to calibrate this spectrum. 
The LDE is here the ratio between the expected number of triggered pixels and the number of scintillation photons. The latter should, if applicable, again be corrected for light quenching.

We can compare the result of this method with the value quoted above, obtained from the energy resolution. In this case we use the data obtained with the spectrometer set to its lowest value ($250$~keV). This set point is corrected for energy loss in the trigger module before it is used to calibrate the experimental spectrum. Using this calibration, the interpeak distance is translated to the energy deposition per pixel which using the light yield, corrected for quenching using $kB = 0.123 $ mm/MeV (Fig.~\ref{fig:EffectBirksLiterature}), gives a LDE of $0.955(16) \%$. This is in remarkably good agreement with the previously obtained value from the energy resolution.

\subsubsection{Simulated light detection efficiency} \label{sec:simulation_point}

\begin{figure}
    \centering
    \includegraphics[width=1\linewidth]{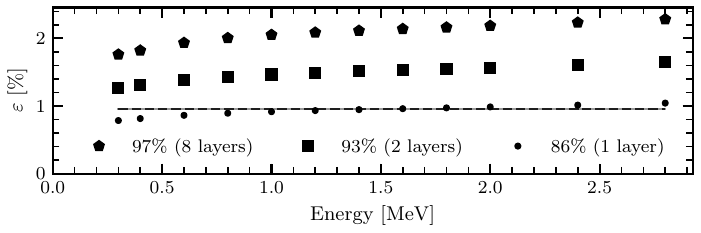}
    \caption{The LDE obtained from the Geant4 simulations for different values of the reflectivity compared to the experimental estimates, indicated by the horizontal lines, introduced in Sec.~\ref{sec:light_collection_eff_InESS}. The reflectivity is compared to the corresponding number of Teflon tape layers reported in Ref.~\cite{Janecek2012}. The roughness parameter $\alpha$ was fixed at $45^{\circ}$ for all simulations.}
    \label{fig:simulated_light_collection_InESS} 
\end{figure}

The light collection estimates discussed above can be compared to the result from simulations for electrons emitted from the exit of the spectrometer, modeled as a circle (\O$1$~mm), in a cone with apex $3.2^{\circ}$. The initial electron energy has a Gaussian distribution around the set point, with $\Delta E/E = 1\%$ at FWHM \cite{Marquet2015}.
In addition to the scintillator, the simulated geometry also includes the trigger module and detector support structures. \\
As indicated above, the scintillator is not polished and is wrapped in Teflon. The ground, back painted model with Lambertian reflection (\ref{app:simulations_G4}) is then the most realistic one to perform the optical simulations in Geant4. These simulations use a constant PDE, $\varepsilon_{\rm PD}=0.391$ (Table~\ref{tab:PDE}).
An improved agreement with the experimental estimates was observed for a higher value of the detector roughness parameter (\ref{app:simulations_G4}), $\alpha >35^{\circ}$. This is expected for this unpolished and thus rough surface. Even closer agreement with the experimental estimates is obtained when the reflectivity probability is reduced to $86\%$, as illustrated in Fig.~\ref{fig:simulated_light_collection_InESS}. According to Ref.~\cite{Janecek2012} this reflectivity corresponds to a scintillator wrapped with only one layer of Teflon tape. Thus, these simulations suggest that the LCE can be strongly enhanced by being careful not to stretch out the tape and using more layers. 
Figure~\ref{fig:simulated_light_collection_InESS} also shows that the simulated detection efficiency increases for higher electron energies. This higher detection efficiency originates from the, on average, deeper implantation within the scintillator, closer to the SiPM. 

\subsection{bSTILED}

The bSTILED (b: Search for Tensor Interactions in nucLear bEta Decay) project aims to extract the Fierz interference term from the most precise measurement of the $^{6}$He decay spectrum. A first experiment used a $25$~keV pulsed $^{6}$He beam from the GANIL-SPIRAL1 target-source system, implanting the ions during a few seconds into the surface of a fixed YAP:Ce scintillator \cite{Kanafani2023}. A movable detector is then placed in contact with the fixed one, enclosing the activity source for a measurement period of typically $12$~s. 
\begin{figure}
    \centering
    \includegraphics[width=\linewidth]{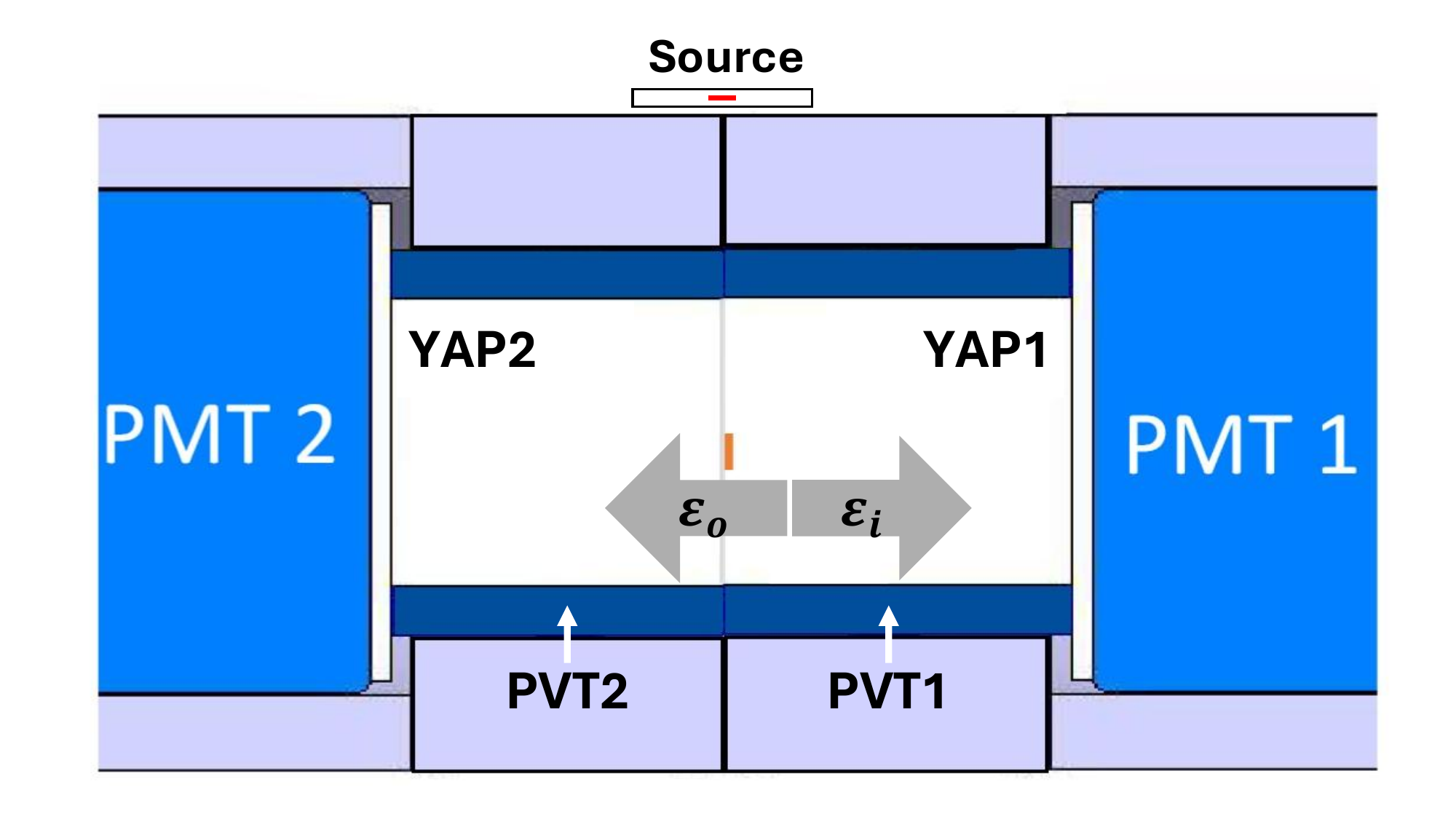}
    \caption{Scheme of the detection setup used in the low-energy bSTILED experiment. The setup consists of two cylindrical (\O$30\times 30$ mm$^2$) YAP:Ce scintillator (white) surrounded by a PVT EJ-204 scintillator (dark blue) which is wrapped in a layer of Tyvek. The location of the $^{6}$He deposition is indicated in orange and the position of the sources for the offline determination of the detector response is indicated at the top. In this setup, a PMT collects scintillation light from the scintillators directly coupled to it, with efficiency $\varepsilon_{i}$, or from the opposite scintillator, with efficiency $\varepsilon_{o}$, as indicated by the gray arrows.} 
    \label{fig:schematic_setup_bSTILED}
\end{figure}

The set-up (Fig.~\ref{fig:schematic_setup_bSTILED}) consists of two cylindrical (\O$30\times~30$ mm$^2$) YAP:Ce scintillator surrounded by a PVT EJ-204 scintillator which is wrapped in a layer of Tyvek. A YAP:Ce and PVT are combined in a phoswich assembly and read out by a single Hamamatsu R7723 PMT.
A pulse shape analysis technique is implemented to discriminate signals originating from the YAP:Ce or from the PVT, based on their different time responses.
The $25$~keV $^{6}$He$^+$ beam is implanted at a depth of $130$~nm in Det.~1 as indicated  in Fig.~\ref{fig:schematic_setup_bSTILED}.
No light reflector was used on the interface between the scintillators, ensuring that no dead layer affects the electron energy measurements. Due to the optical coupling between the two scintillator assemblies, a fraction of the light emitted by one scintillator crosses the interface and is detected by the PMT of the opposite assembly. For a given PMT, we define the LCE, $\varepsilon_i$ for the scintillation light emitted by the scintillator directly coupled to that PMT, and $\varepsilon_0$ when it was emitted by the opposite scintillator (Fig.~\ref{fig:schematic_setup_bSTILED}).

\subsubsection{Light collection efficiency}
\label{sec:light_collection_eff_bSTILED}

In the bSTILED experiment the response function of the detectors was characterized offline using gamma-sources \cite{Kanafani2023PhD}. The photopeaks ranging from 59.54 keV ($^{241}$Am) to 2614.5 keV ($^{208}$Tl) were fitted with Gaussian curves and the resulting plot of the FWHM as a function of energy was subsequently fitted with Eq.~\eqref{eq:res_exp}. The $\alpha$ and $\gamma$ parameters were found to be negligible.
The total LDE is taken again as the product between the LCE and the PDE,
$\varepsilon_{\rm LD} = \varepsilon_{\rm LC}\cdot\varepsilon_{\rm PD}$. From Table~\ref{tab:PDE} we have $\varepsilon_{\rm PD} = 0.243$ and thus Eq.~\eqref{eq:res_counts} leads to an expected LCE of $47.7\%$ for one detector and $49.1\%$ for the other.
The resolution was also determined online, together with the energy calibration, using an iterative procedure which makes successive comparisons between the simulated and experimental histograms \cite{Kanafani2025}. The comparison is dominated by events which deposit their full energy in one detector and neglects the $\alpha$ and $\gamma$ parameters.
Under such conditions, and using again $\varepsilon_{\rm PD} = 0.243$ (Table~\ref{tab:PDE})
the measured $\beta$ parameters correspond a LCE of $44.2\%$ for one detector and $39.0\%$ for the other.
The resolution, and correspondingly, the LCE determined using $\gamma$-sources differs from the online analysis using $\beta$-decays. This is not unexpected since the optical coupling of the detection set-up changed between both measurements. But there is also a more fundamental reason to expect a difference between the two. Compared to $\gamma$-rays, the interactions of $\beta$-particles typically occur in a smaller volume of the detector due to their shorter range. The size of this difference is determined by the detector geometry and is discussed in Sec.~\ref{sec:simulation_geometry}.

We compared the light collection estimates discussed above to the result from simulations for particles emitted isotropically over a $4\pi$ sr solid angle from a point source located $130$~nm inside Det.~1 and at radius $0$~mm, which is the center of the implantation position for the $^{6}$He$^+$ beam at GANIL.
The scintillators were polished and the PVT surrounding the YAP:Ce crystals were wrapped in a layer of Tyvek. This makes the ground, back painted model with specular spike reflection the most realistic one. 
It turned out to be impossible to reproduce the experimental energy resolution using the, to our knowledge, only available literature value for the wavelength-dependent absorption length \cite{Baccaro1998}. A LCE of only $\varepsilon_{\rm LC}(661~\text{keV})=15.7\%$ was found,
which is about three times smaller than expected. 
Although quite inaccurate, the datasheet provided by the supplier indicates an absorption length that can be approximated by a step function, which varies from $0$ to $100$~cm at $330$~nm. This function, resulting in $\varepsilon_{\rm LC}(661~\text{keV})=48.8\%$, was used for the analysis.

The simulated fraction of light shared between both detectors due to the optical cross talk can also be compared to the experiment. The light collected by a PMT from the opposite scintillator relative to the light from the scintillator directly coupled to that PMT was determined to be $18 - 20 \%$ \cite{Kanafani2023PhD}.
This can be compared to the simulated ratio for $\varepsilon_o/\varepsilon_i$ and both were found to be consistent (see below).
Remarkably, good agreement was also found when using the wavelength-dependent absorption length from Ref.~\cite{Baccaro1998} which yields a value of about 21\%.

\subsubsection{Light collection maps}
\label{sec:simulation_geometry}

The difference in interaction positions inside the detector, for $\gamma$-rays and $\beta$-particles, can be expected to have an important effect in scintillators for which either the light attenuation length is comparable to the scintillator size or for irregular detector shapes. We tested its importance for the YAP:Ce scintillator in the bSTILED experiment. 

\begin{figure}
    \centering
    \includegraphics[width=\linewidth]{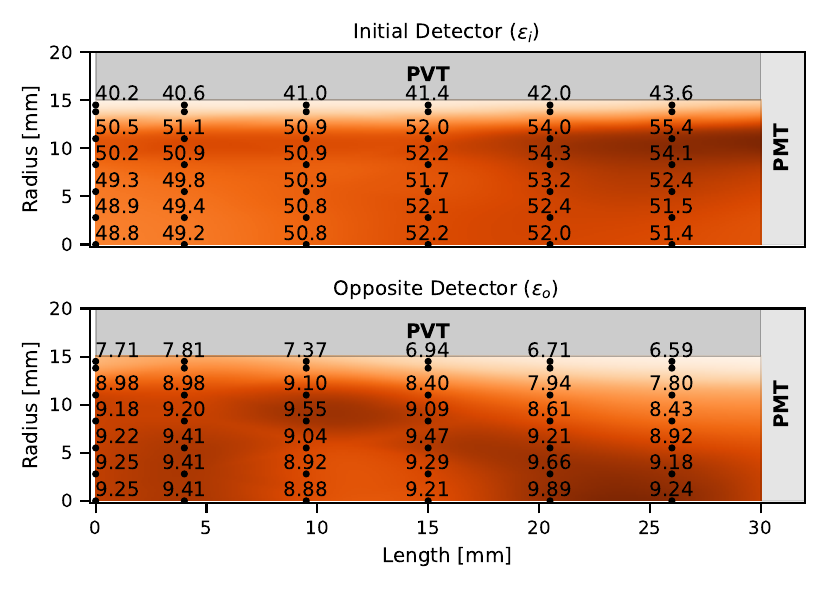} 
    \caption{Simulated LCE, $\varepsilon_{\rm LC}$ (in \%) for electrons of $500$~keV emitted at the positions indicated by the dots, for: (top) light emitted by the scintillator originally coupled to the PMT, $\varepsilon_i$;
    (bottom) light emitted by the opposite scintillator, $\varepsilon_o$.
    The figure is limited to positive radii due to the cylindrical symmetry of the system. The color map is determined by an interpolation and only serves as a guide to the eye.}
    \label{fig:light_collection_map}
\end{figure}

The position dependence of the LCE has been determined by simulating isotropically emitted mono-energetic electrons at different positions in the crystal. 
An example of the resulting light collection map is shown in Fig.~\ref{fig:light_collection_map}.
Given the accurate pulse shape discrimination in the experimental spectra, only electrons which deposit their full energy in one of both YAP:Ce crystals, were selected. 
For each point, $50~000$ electrons were generated. 
The map shows regions with varying collection efficiencies which are mainly induced by the surrounding PVT detector and the optical cross-talk between both scintillator assemblies.

\begin{figure}
    \centering
    \includegraphics[width=\linewidth]{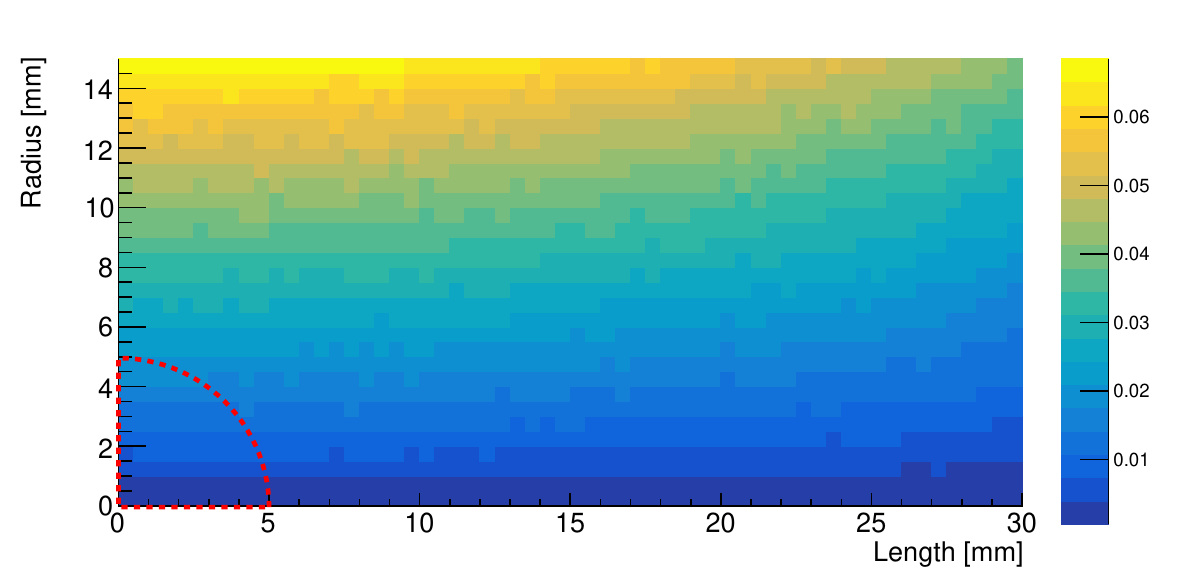}
    \caption{Simulated interaction probability (in \%) for a $\gamma$-ray of $E_\gamma=661$~keV emitted by a calibration source installed just outside of the detection set-up (Fig.~\ref{fig:schematic_setup_bSTILED}).
    See text for details.}
    \label{fig:gamma_interaction_map}
\end{figure}

The simulated interaction positions for $\gamma$-rays of $E_\gamma=661$~keV emitted by a calibration source installed just outside of the detection set-up as used to determine the detector response \cite{Kanafani2023PhD} are shown in Fig.~\ref{fig:gamma_interaction_map}.
Measurements like this have initially been used to study the detector response \cite{Kanafani2023PhD}. However, for the $\beta$-particles emitted at the location of the $^{6}$He deposition (length $130$~nm and radius $0$~mm) with an average range of only about $5$~mm for a $3.5$~MeV electron (which is the endpoint of the $^6$He spectrum), most interactions occur at a different position inside the crystal as indicated by the dotted red circle.
As a result, the LCE for both measurements is different (as shown in Fig.~\ref{fig:light_collection_map}) thus making the detector response measured with the $\gamma$ sources less reliable for $\beta$-decay studies.
More precisely, the resulting simulated LCE for $661$~keV $\gamma$ rays is $\epsilon^{\gamma}_{\rm LC}=51.4\%$. This is $2.6\%$ higher than for $661$~keV electrons emitted at the location of the $^6$He deposition (Sec.~\ref{sec:light_collection_eff_bSTILED}). This higher LCE can be understood as the average after the convolution of Figs.~\ref{fig:light_collection_map} and \ref{fig:gamma_interaction_map}.

\begin{figure}
    \centering
    \includegraphics[width=\linewidth]{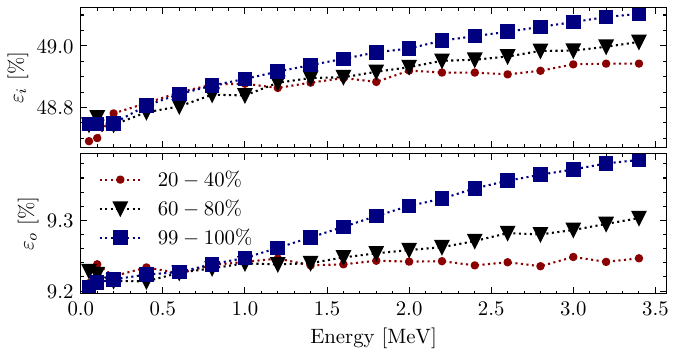}
    \caption{LDE for scintillation light produced by electrons emitted at the location of the $^6$He deposition ($L=130$~nm and $R=0$~mm). Top: light collected by the PMT mounted on the scintillator, $\varepsilon_{i}$; Bottom: light collected by the
    opposite PMT, $\varepsilon_{o}$. The different curves correspond to the indicated fractions of the total electron energy deposited in the detector.}
    \label{fig:bSTILED_edep_collection}
\end{figure}

The light collection map in Fig.~\ref{fig:light_collection_map} also shows that electrons emitted deeper in the scintillator have an increased LCE. The same occurs for electrons with higher energy as, on average, they travel deeper inside the scintillator. This effect was already described for the detector in the InESS experiment, as was shown in Fig.~\ref{fig:simulated_light_collection_InESS}.
Figure \ref{fig:bSTILED_edep_collection} shows the LCE for electrons of different energies emitted at the center of the $^{6}$He implantation region.
Each panel shows the curves for three different intervals of the energy deposited by an electron in a single scintillator. Whereas some electrons deposit their full energy, $99-100\%$, in a single detector, others are backscattered and their deposited energy is shared between both scintillators. The LCE increases for events that deposit their full energy in the scintillator coupled to the PMT whereas this effect is much less pronounced for events that are backscattered and thus deposit $20-40\%$ of their total energy. This is expected as these backscattered electrons do not penetrate as deep in the detector.

\subsection{Cherenkov in Geant4}

When introducing the Cherenkov effect in Sec.~\ref{sec:Cherenkov}, we assumed an equal
LCE for all photons, irrespective of their origin. However, Cherenkov and scintillation photons have a different directionality. The first are emitted in a cone along the electron trajectory whereas the latter are emitted isotropically. Moreover, the LCE also depends on the electron energy due to the range within the detector material, as illustrated in Figs.~\ref{fig:simulated_light_collection_InESS} and \ref{fig:bSTILED_edep_collection}. In summary, some photons have a more direct route towards the PMT thereby potentially modifying the nonlinearities shown in Fig.~\ref{fig:Cherenkov_nonlinearity_example}.

The production of Cherenkov photons is natively implemented into the optical models of Geant4. To incorporate the associated nonlinearity, it is sufficient to specify the refractive index of the detector material. However,  the refractive index should be defined only over the spectral range corresponding to the PMT (Fig.~\ref{fig:ES_overview}) to avoid overestimating the effect.
A previous study \cite{Trigila2022} reported that modifying the maximal variation in electron velocity per step from $0.01\%$ to $50\%$ influences the average step length and the mean number of Cherenkov photons. We checked this observation in the simplest possible model, namely with mono-energetic electrons emitted isotropically from the center of a sphere made of a specific material which is large enough to stop all electrons ($R = 5$~cm) and saving the generated number of Cherenkov photons. In this configuration, the only escaping particles are bremsstrahlung photons. When comparing the average number of simulated photons with the result of Eq.~\eqref{eq:theory_cherenkov_number} we found excellent agreement and thus do not observe the previously reported deviations when varying the maximal electron velocity per step.

\begin{figure}
    \centering
    \includegraphics[width=\linewidth]{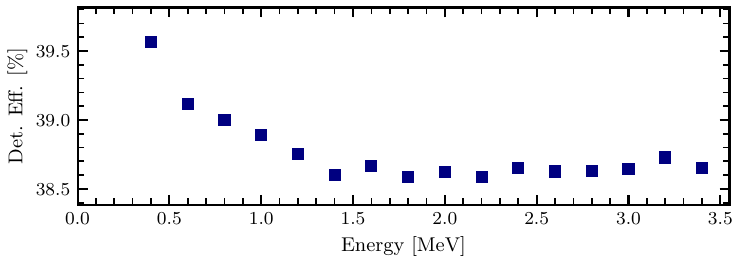} 
    \caption{The detection efficiency of Cherenkov photons at different electron energies for the bSTILED setup. At lower energies, the Cherenkov photons have an increased detection efficiency. This effect is mitigated at higher energies due to the irregular trajectory of the electron in the detector.
    }
    \label{fig:CherenkovDirectionality}
\end{figure}

The efficiency to detect Cherenkov photons emitted by electrons of different energies in the bSTILED setup is shown in Fig.~\ref{fig:CherenkovDirectionality}.
Electrons have again been emitted at the center of the implantation position for the $^{6}$He beam but directly towards the PMT. The detection efficiency is then calculated for electrons that are not backscattered, but have been fully stopped in one of both scintillators. The Cherenkov photons, being emitted in a cone, are seen to have an increased detection efficiency at lower energies. At higher energies, this effect is reduced due to the electrons irregular trajectory in the detector.
We recall that for scintillation photons, the detection efficiency increases linearly with the electron energy due to the, on average, deeper implantation depth as was shown in Fig.~\ref{fig:bSTILED_edep_collection}. 

\section{Extraction of the Fierz interference term}
\label{sec:MCfits}


Following the presentation of several sources of nonlinear response in a scintillator, we focus here on quantifying the effect for the extraction of the Fierz interference
term from measurements of beta-energy spectra.
The procedure was inspired by previous work \cite{Gonzlez-Alonso2016} and consisted in analyzing
pseudo-experimental spectra generated by a MC routine. 
The spectra were generated following the shape of the allowed phase space, without including corrections, $S_{\rm SM} (W) = pWq^2$, where $p = \sqrt{W^2 + 1}$ and $W = E_e/m_e + 1$ are respectively the momentum and total energy of the $\beta$ particle and $q = W_0 - W$ is the momentum of the neutrino. Each spectrum contained $10^8$ events and was then fitted between $5\%$ and $95\%$ of the kinetic energy range with a function of the form
\begin{equation} \label{eq:Fierzfit}
    S (W) = A \left( 1 + \frac{\gamma}{W} b \right) S_{\rm SM}(W)
\end{equation}
with $\gamma = \sqrt{1-(\alpha Z)^2}$. The overall normalization $A$ and the Fierz term $b$ are free parameters. The finite energy resolution of detectors was neglected both in the pseudo-experimental spectra and in the fit function.

\begin{figure}
    \centering
    \includegraphics[width=\linewidth]{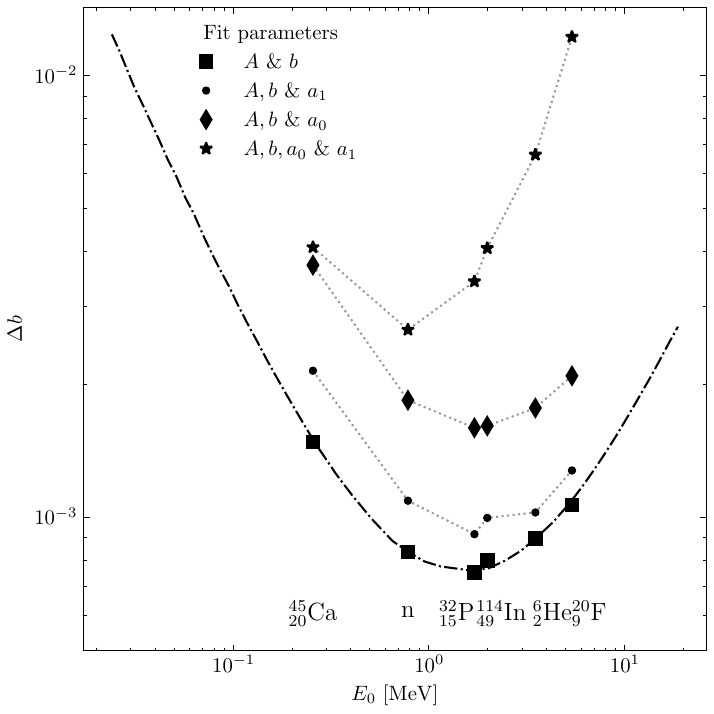}
    \caption{Uncertainties on the Fierz interference term (at $1\sigma$) obtained from fits of spectra with $10^8$ events using different sets of free fit parameters. The spectra were fitted between $5\%$ and $95\%$ of the beta-particle kinetic energy range. Simulations were performed for isotopes indicated at the bottom of the figure. The black, dash-dotted line is the result presented in Ref.~\cite{Gonzlez-Alonso2016}. The simulations are repeated for different sets of free fit parameters (see text for details) which shows the corresponding evolution of the uncertainty on the Fierz term, $b$. }
    \label{fig:bFsensitivity}
\end{figure}

The experimental sensitivity was then determined from the $1\sigma$ statistical uncertainty, $\Delta b$, on the Fierz term obtained from these ﬁts. 
Figure \ref{fig:bFsensitivity} shows the dependence of $\Delta b$ as a function of the end-point energy, as already reported in Ref.~\cite{Gonzlez-Alonso2016}, with some relevant isotopes for spectrum shape measurements indicated. 
This shows that the sensitivity of $^{114}$In is slightly smaller than what could be deduced from the similar plot in Ref.~\cite{Gonzlez-Alonso2016} based only on the $Q$-value, due to the isotope's higher $Z$ which reduces the factor $\gamma$ in Eq.~\eqref{eq:Fierzfit}.

In practice, an experimental energy spectrum is obtained in ADC units, $C$, so assuming a linear calibration we have $E = a_0 + a_1 C$. As suggested elsewhere \cite{Huyan2018}, it is difficult to reach the targeted precision in an independent external
calibration and the calibration parameters could be let free in the fit with a so-called "auto-calibration method". Figure~\ref{fig:bFsensitivity} shows the evolution of the uncertainty on the Fierz term when including $a_0$ and $a_1$ as free parameters. It is clearly more advantageous to only include $a_1$ as a free fit parameter since leaving (also) $a_0$ free strongly increase the statistical uncertainty due to correlations.

This kind of fits can also be used to quantify the systematic shift associated to the sources of nonlinearity in the detector response. This was done by looking at the value of $b$ when including the effect of the nonlinearities in the spectrum generated by the MC routine but not at all or partially in the fit. 
In the following, the quoted shifts in the values of $b$ correspond to fits where $A$, $b$, and $a_1$ were free parameters for the most relevant isotopes indicated in Fig.~\ref{fig:bFsensitivity}. It is important to note that the auto-calibration method can mask the effect of the nonlinearities being studied. The fitted value for $a_1$ then partially compensates the unaccounted-for effect resulting in an acceptable fit but with an erroneous value for $b$.

\subsection{Light quenching}

The effect of light quenching (Sec.~\ref{sec:Birks}) strongly depends on the scintillator material. The light output of YAP:Ce is generally accepted to be linear above $10$~keV \cite{Mengesha1998, Kapusta1999, Moszyski2016}, whereas it will strongly modify the experimental energy spectrum for a PVT detector as shown in Fig.~\ref{fig:EffectBirksLiterature}.
For example, correcting the theoretical spectrum with $kB = 0.123$~mm/MeV for an experimental spectrum obtained with $kB=0.143$~mm/MeV results in a fitted value for $b$ which is shifted by about $1\%$ on average (Table~\ref{tab:bF_Birksnonlinearity}) with all fits still having an acceptable $\chi^2$. 
In other words, for PVT one needs to know the value for $kB$ with a relative accuracy smaller than $1.5\%$ to reach a sub permille precision on $b$.
The shift depends modestly on the end point energy because quenching is most important for low-energy electrons (Fig.~\ref{fig:EffectBirksLiterature}).

\begin{table}
    \centering
    \caption{Systematic shift (or error) on $b$ related to the possible sources of nonlinearity discussed in the text: i) quenching of the scintillation light; ii) the presence of Cherenkov radiation and how it is affected by a shift of the spectral sensitivity of the readout device; and iii) the energy response of a SiPM. The first and second column indicate the effect and the investigated parameter. The other entries list the systematic shift, $\delta b_F$, of the Fierz term (in \%). See text for details.}
    \label{tab:bF_Birksnonlinearity}
    \begin{tabular}{l lcccc}
    \hline \hline
    & & \multicolumn{4}{c}{$\delta b_{sys} [\%]$} \\
      &  & $n$ & $^{114}$In & $^{6}$He & $^{20}$F   \\
    \hline
    Quenching & \multicolumn{1}{c}{$kB$} & 1.71 & 1.04 & 0.80 &  0.76 \\
    Cherenkov  & bSTILED & -0.46 & -0.38 & -0.31 & -0.19 \\
	           & ~~spec. shift & 0.00 & 0.02 & 0.04 & 0.11 \\
	& InESS & -4.25 & -4.34 & -4.11 & -4.05 \\
	& ~~spec. shift & -0.13 & -0.12 & -0.09 & -0.04 \\
    SiPM  & & 1.83  & 4.39 & 8.28 & 15.52 \\
    ~~~~$P_{cross}$ & $2\%$ & 0.02 & 0.06 & 0.11 & 0.19 \\
                    & $4\%$ & -0.05 & -0.12 & -0.21 & -0.39 \\
    ~~~~$E_{pixel}$ & $9$~keV & -0.22 & -0.50 & -0.92 & -1.68 \\
                    & $11$~keV & 0.16 & 0.36 & 0.67 & 1.22 \\
    \hline \hline
    \end{tabular}
\end{table}

\subsection{Cherenkov}

The shift on $b$ induced by the nonlinearity due to Cherenkov photons is much smaller than for the effect of light quenching as could already be expected by comparing the vertical axis of Figs.~\ref{fig:EffectBirksLiterature} and \ref{fig:Cherenkov_nonlinearity_example}.
Nevertheless, as shown in line 4 of Table~\ref{tab:bF_Birksnonlinearity}, when the effect is overlooked in the fit (Eq.~\eqref{eq:Fierzfit}), the shift of $b$ for the InESS experiment using a PVT detector coupled to a SiPM is up to about 4\%. For the bSTILED experiment (line 2), the systematic shift would be limited to a few permille due to the larger density of the YAP:Ce scintillator and the larger light yield.  
%
Fortunately, this effect can accurately be taken into account using the analytical description discussed previously, Eq.~\eqref{eq:theory_cherenkov_number}. The remaining uncertainty is then limited to the calculation of the average PDE, as reported in Table~\ref{tab:PDE}. This calculation requires the scintillation emission spectrum and the spectral sensitivity, PDE, of the readout device (Fig.~\ref{fig:ES_overview}). Both are, however, often only given as a graph in the respective data sheets. To get a conservative estimate on the possible associated deviation, we investigated the influence of shifting the scintillation spectrum and the PDE spectrum by $10$~nm. 
The strongest effect is seen when shifting the PDE spectrum since such a shift modifies the spectral sensitivity window of the detector system and thus its sensitivity for Cherenkov photons.
However, as shown in lines 3 and 5 of Table~\ref{tab:bF_Birksnonlinearity}, even the largest shifts are comparable to the statistical uncertainty for a spectrum with $10^8$ counts (Fig.~\ref{fig:bFsensitivity}).
A similar $10$~nm shift of the scintillation emission spectrum mainly modifies the detection efficiency for the scintillation spectrum. The resulting systematic shift on $b$ is of the order of a few $\mathcal{O}(10^{-4})$.
In general, we can conclude that, when not overlooked, the systematic error or shift on $b$ associated to Cherenkov photons can be controlled to reach the permille level on $b$.

\subsection{SiPM nonlinearity}

In principle, improving the LCE results in more reliable experimental spectra due to the improved energy resolution. However, when using a SiPM, its intrinsic nonlinearity increases when more pixels are triggered (Fig.~\ref{fig:SiPM_non_linearity}). For the same reason, the importance of this correction drastically increases for beta transitions with higher endpoint energies. The effect of this correction will strongly depend on the specific detection set-up. Here, we illustrate its importance for the InESS project using a SiPM with $14~400$ pixels and an expected cross-talk probability of $P_{cross} = 3\%$ \cite{Vanlangendonck2023, HamamatsuTechNotes}. Using a LDE of $1\%$ (Sec.~\ref{sec:light_collection_eff_InESS}), the estimated energy per pixel triggered is about $E_{pixel}= 10$~keV, due to the large mismatch between the area of the SiPM and that of the scintillator. \\
Having a value for the LDE it is possible to calculate the correction discussed in Sec.~\ref{sec:readout} using the distribution calculated from Eq.~\eqref{eq:SiPM_nonlinearity_distribution} and a Binomial distribution to model cross-talk by calculating their mean. Line 6 in Table~\ref{tab:bF_Birksnonlinearity} lists the shift on $b$ when the effect is not included in the fits. As expected, the systematic error scales strongly with the endpoint energy ranging from about 2\% for the neutron to about 16\% for $^{20}$F. Note that using the transfer function in Eq.~\eqref{eq:NmbPixelsSiPM} rather than the here-presented description, ``Eq.~\eqref{eq:SiPM_nonlinearity_distribution} + cross-talk'', gives an effect twice smaller.

Finally, we can investigate the sensitivity of these results to the parameters in the model, namely the energy per pixel determined by the LDE and the cross-talk probability. To get an order of magnitude estimate (last four lines in Table~\ref{tab:bF_Birksnonlinearity}) we assumed a $1$~keV uncertainty on the energy per pixel, which corresponds to a $\pm10\%$ relative error on the LDE, and a $1\%$ uncertainty on the cross-talk probability. In these fits the MC routine still generates a spectrum using  $P_{cross} = 3\%$ and $E_{pixel} = 10$~keV but the fits include the values listed in Table~\ref{tab:bF_Birksnonlinearity}.
As expected, the effect again scales with the endpoint energy. Moreover, the uncertainty on these parameters results in an asymmetric error on $b$. When the parameters in the model over-correct the fit function, using a higher $P_{cross}$ or LDE than the actual values in the (pseudo-)experimental data, the systematic error $\delta b$ is larger than when the correction is underestimated.

To summarize, the precision needed on the LDE and the cross-talk probability to reach a permille level precision on $b$ strongly depends on the end point energy. But, in general, the uncertainty on the LDE should not be more than a few percent and for the cross-talk probability limited to about one percent.

\subsection{Light collection}

The LCE is strongly experiment specific as it depends on the detector geometry, the detector wrapping, and the optical coupling. Its absolute value can be estimated by the energy resolution, as explained in Sec.~\ref{sec:examples} which, furthermore, allows benchmarking optical simulations. This kind of simulations provide an excellent tool to study the energy dependency of the LCE or variations throughout the detector volume, as shown in Figs.~\ref{fig:simulated_light_collection_InESS}, ~\ref{fig:light_collection_map}, and \ref{fig:bSTILED_edep_collection}. We can use the simulations presented in Fig.~\ref{fig:bSTILED_edep_collection} to get an estimate on the size of this effect when studying the $^{6}$He decay. For events where the total electron energy is deposited in one detector, indicated by $99-100\%$, the resulting systematic error on $b$ is $-1.1\%$ making the energy dependence of the light collection a non-negligible source of nonlinearity for the bSTILED experiment. 

\section{Conclusion}

In this work we studied three sources of nonlinearities in the response of scintillation detectors and determined their impact for the extraction of the Fierz interference term in precision measurements of beta-energy spectra. 
As expected, for detectors where light quenching is important, like PVT, this is the dominant source of nonlinear response. For a PVT detector, one needs to know the value of $kB$ in the Birks formula (Eq.~\ref{eq:Birks}) with an accuracy smaller than $1.5\%$ to get sub permille precision on $b$. It is important to note that the inconsistencies between the literature values for this commonly used material are larger than $10\%$.

Cherenkov radiation is an often overlooked source of detector nonlinearity, which is more important for detectors with a small light output and low density. Moreover, its importance will depend on the PDE of the readout device. When not corrected for, it results in a deviation on $b$ by a few permille for a high-density detector with relatively high light output such as YAP:Ce crystals and up to a few percent for PVT.

The detector response of a SiPM has an intrinsic energy dependence due to the finite number of pixels working in Geiger mode. We presented a formula to take this into account by calculating the number of pixels triggered given the number of incident photons and the total number of pixels. This formula extends the scope of the input-output transfer function presented in the literature. 
This intrinsic nonlinearity increases for higher light-collection efficiencies and higher energies. Nevertheless, sufficient light should be collected to guarantee a reliable determination of the detector response.

Finally, a method has been presented to estimate the LCE, which turns out to be a useful tool for a reliable inclusion of optical tracking in MC simulations. We showed, for example, how they allow quantifying the difference in detector response between a $\gamma$- and a $\beta$-source.

These findings highlight the necessity of accurately accounting for such nonlinearities to achieve permille precision in the extraction of the Fierz interference term. The presented methods and simulations provide a robust foundation for their inclusion in the analysis and planning of future experiments aiming at
precision measurements of the beta-energy spectrum with a scintillation detector.

\section*{Acknowledgements}
This work was supported by the Research Fund Flanders (FWO) through the IRI projects I001323N, I002219N, I002619N, and I002929N, by the KU Leuven project C14/22/104,
and by the Agence Nationale de la Recherche under grant ANR-20-CE31-0007-01 (bSTILED) and ANR-18-CE31-0004-02 (WISArD). 

\appendix

\section{Optical simulations in Geant4} \label{app:simulations_G4}

Before optical tracking can be included in Geant4 simulations, the user should provide the necessary input parameters such as the material properties. The most important parameters are the refractive index\footnote{When an optical photon reaches the border of a volume where the refractive index has not been defined, it is not tracked further (killed).} and the scintillation light yield. Additionally, to obtain acceptable performance, it is often necessary to evaluate the boundaries between the different components of the detection geometry. At first, this is done by including the necessary extra volumes for a realistic scintillator geometry as for example the optical grease between the scintillator and its corresponding photomultiplier. 
Thereafter, appropriate boundary models should be used to include the detector wrapping. For some materials, it is possible to use look-up tables determined from dedicated experiments, i.e., BGO (LNBL look-up table - LUT) \cite{Janecek2010} and L(Y)SO (DAVIS) \cite{Stockhoff2017}. However, in general, such information is not available and the properties of the boundary are determined by providing the necessary parameters in the unified model. 

\begin{figure}
    \centering
    \includegraphics[width=\linewidth]{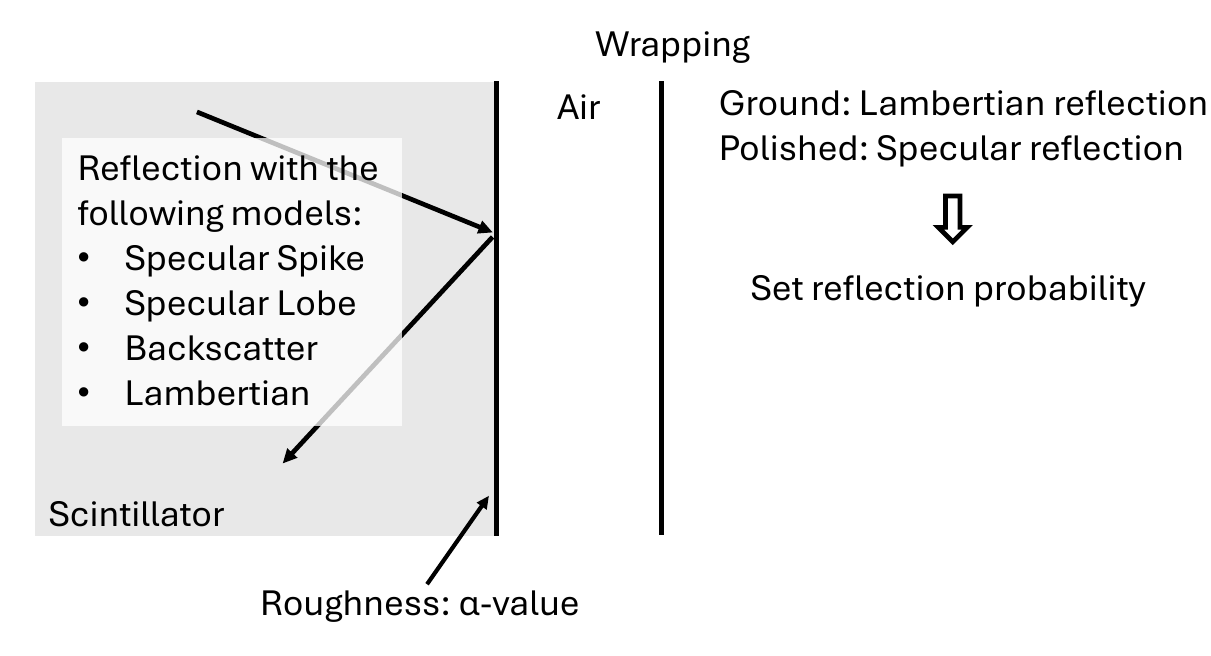}
    \caption{Summary of the parameters that can be used to optimize the reflection of optical photons on a back painted boundary in the unified model. Lambertian reflection models the reflection on a perfectly diffuse surface and is thus a good estimate for rough surfaces.}
    \label{fig:summary_G4models}
\end{figure}

The unified model provides four boundary types depending on the reflectivity of the wrapper surrounding the scintillator (a polished or a rough, called ground, reflector) and on the presence of an air layer in-between this wrapper and the scintillator (front or back painted) \cite{Kandemir2018}. 
With the appropriate model selected, the simulations can further be optimized by tuning several parameters as summarized for the back painted model in Fig.~\ref{fig:summary_G4models}. 
The roughness of the scintillator surface is modeled by subdividing the surface into numerous micro-facets with the angle $\alpha$ defining the average angle between the normal to the surface and the micro-facets. Once the normal of the micro-facet is sampled, the reflection type is determined according to the defined probabilities for specular spike, specular lobe, backscatter, or Lambertian reflection. An illustration of these models is given, for example, in Ref.~\cite{Kandemir2018}. 
The total reflectivity of the scintillation wrapping can be established in dedicated measurements of the reflectivity probability \cite{Janecek2012, Janecek2008}. 

Finally, to avoid unrealistically long tracks, the wavelength-dependent absorption length of the scintillator can be included. This parameter becomes especially important when its dimensions are comparable to the size of the detector. Its value should again be determined from dedicated experiments \cite{Baccaro1998}. As mentioned before, the accuracy of the selected parameters can be evaluated by comparing the simulated results with the LCE estimates.

The most appropriate approach to modify the detection geometry is sometimes, \textit{a priori}, unclear. For the simulations of the bSTILED geometry \cite{Kanafani2022} discussed above, it works well to modify the geometry by introducing a small gap, of about \SI{1}{\micro m}, between the two touching YAP:Ce crystals. If not, both are treated as a single detector when tracking the optical photons, resulting in an unrealistically large amount of optical cross-talk between the detectors. Looking at Fig.~\ref{fig:summary_G4models}, however, a polished back painted boundary between both detectors might seem to be a good alternative. The superior performance of the former approach is then confirmed by comparing to the experimental estimates presented in this work.

\bibliographystyle{unsrt} 
\bibliography{references}

\end{document}